\documentclass[preprint,aps,preprintnumbers,longbibliography,longbibliography]{revtex4-1}

\usepackage[latin9]{inputenc}
\usepackage{amsmath}
\usepackage{amssymb}
\usepackage{graphicx}
\usepackage{esint}
\usepackage[scr=boondoxo, scrscaled=1.05]{mathalfa}
\allowdisplaybreaks
\usepackage{setspace}
\makeatletter

\pdfpageheight\paperheight
\pdfpagewidth\paperwidth
\usepackage[top=1in,bottom=1in,left=1in,right=1in]{geometry}

\pdfoutput=1

\usepackage{epsfig}
\usepackage{graphics}

\usepackage{mciteplus}
\mciteErrorOnUnknownfalse

\makeatother

\begin{document}

\title{Cosmic shadows of  causation}
\large
\author{Craig  Hogan}
\affiliation{University of Chicago}

\date{\today}
\begin{abstract}
\large
 Cosmic structure on the largest scales  preserves the pattern laid down by quantum fluctuations of gravity in the early universe on scales comparable to  inflationary horizons.
It is proposed here that fluctuations create physical correlations only within finite regions  enclosed by causal diamonds, like entanglement in other quantum systems. 
Conformal geometry  is used to calculate a range of angular separation with  no causal correlation.  Correlations of the cosmic microwave background  in this ``causal shadow'' are  measured to be three to four orders of magnitude smaller than expected in standard inflation models. It is suggested that this  measured symmetry of the primordial pattern  signifies a new  symmetry of  quantum gravity associated with causal horizons.
\end{abstract}
\maketitle

\large
{\it But what was there in the beginning? A physical law, mathematics, symmetry? In the beginning was symmetry!} 
\begin{flushright}--- Werner Heisenberg \cite{Heisenberg1971}
\end{flushright}

\vspace{.4in}

About thirty years ago, the Cosmic Background Explorer satellite  ({\sl COBE}) delivered the first maps of the  primordial pattern in the cosmic microwave background (CMB). They provided spectacular evidence of an amazing fact: the very early universe created coherent primordial gravitational structure on the very largest scales.  Correlations measured in later CMB experiments with better precision and angular resolution now confirm a primordial origin for gravitational perturbations that generate the entire  vast large-scale structure of the cosmic web, and the eventual gravitational binding of all astronomical systems.  

Cosmologists agree that  this structure was  laid down in a  pattern determined by quantum fluctuations of gravitational potential during an early era of accelerating expansion. During this inflationary era,   quantum fluctuations on a microscopic scale coherently distorted   space-time curvature.  According to inflation theory,  distortions were frozen into perturbations of gravitational potential, and then amplified to enormous size, by the rapidly accelerating expansion.  The pattern now measured on the sky and in large-scale structure is not random, uncorrelated noise, but a superposition of these coherent distortions. Over a large range of scales, the measured correlations of the noise agrees with the scaling expected from the inflationary expansion. 

Another remarkable fact revealed by {\sl COBE} \cite{Bennett1994,Hinshaw_1996}, and later confirmed more precisely by  the {\sl WMAP} and {\sl Planck} satellite teams\cite{Bennett2003,WMAPanomalies,Planck2016,Akrami:2019bkn}, is that the primordial pattern on the very largest angular scales--- the largest structures of any that we can measure, comparable to the size of the whole observable universe limited by our cosmic horizon---  is strangely uniform.  The uniformity is conspicuous in a simple statistical measure called the angular  correlation function, $C(\Theta)$,  the mean product of temperature in pairs of points separated by angle $\Theta$. Although the fine-scale statistical pattern agrees with theory to high precision, the  correlation  at angular separations  larger than  about sixty degrees is much closer to zero than typically expected  from the same theory that agrees so well with measured structure at smaller scales.  

The absence of large scale correlation has been the subject of much study and debate over the decades since {\sl COBE} first found it\cite{Copi2008}.  The most widely held view is that it is just  accidental: the standard theory predicts many possible outcomes, and it just so happens that the large-scale patch of the universe that we live in is a   rare pocket of uniformity.  If we  picture primordial curvature perturbations as waves on a storm-tossed sea,  we happen to find ourselves in a large, unusually becalmed region, where  the  largest waves we can measure happen to nearly cancel each other.

Let us  consider an alternative possibility, that the large scale uniformity of the CMB sky is a  signature of a new physical symmetry.  An empirical reason to take this view seriously is that as  CMB maps have improved,  the magnitude of the measured large angle  correlation has become smaller. At angular separations near 90 degrees, it appears to be consistent with zero, within the measurement uncertainties from models of Galactic contamination\cite{Hagimoto_2020,hogan2023causal}. 
Another reason to consider this view is that
 anisotropy at large angular separation is dominated by fluctuations at long wavelengths, a regime where  standard field approximations require modification to account for gravitational entanglement with causal structure    \cite{CohenKaplanNelson1999,HollandsWald2004,Stamp_2015}.  It could be that  standard inflation theory fails to predict a true physical symmetry because it does not correctly describe  causal entanglement of fluctuations on scales comparable to inflationary horizons.

In general, physical quantities must be part of the same quantum system to be correlated, and  they must be causally connected.  Since large scale cosmological perturbations arise from a quantum system, it is natural to ask whether they  show a symmetry  connected with causal limits of information propagation during inflation. 
Quantum uncertainty  does not violate causality:  even  ``spooky'' nonlocal quantum information never travels faster than light\cite{Zeilinger1999}.  As far as we know, this well tested constraint extends to  nonlocal coherence of quantum fluctuations on finite intervals of world lines: a  fluctuation  at one place is entangled with fluctuations at other places, but only within bounds of two-way casual communication determined  by the  future and past light cones that bound the interval, which encompass a ``causal diamond'' in four dimensional space-time. 
Physical effects of quantum fluctuations in separate causal diamonds are independent and uncorrelated, because separate causal diamonds encompass separable quantum systems. Based on these principles, let us posit a specific causal boundary of quantum coherence: {\it  all coherent  physical effects entangled with a quantum  fluctuation over a timelike interval lie within its causal diamond.} 

During cosmic inflation,  the accelerating expansion creates horizons around every world line.  Unlike a  black hole horizon,  we can view the system after the horizon disappears at the end of inflation, so we can see what happened. But since a horizon bounds information flow during inflation where perturbations originate, it  also bounds  entanglement, and hence correlation: no information that influences the curvature perturbation at any location can reach it from another location after it crosses the horizon, so {\it  all  of the measured correlation in the late universe  arises from quantum fluctuations within causal diamonds bounded by inflationary horizons.}

To understand the effect of this constraint, it is necessary to  analyze  inflationary causal relationships in  four dimensions. In conformal comoving coordinates, the    metric of any spatially uniform and isotropic universe takes the form
\begin{equation}\label{FLRW}
    ds^2 = a^2(t) [c^2d\eta^2- d\Sigma^2],
\end{equation}
where $a(t)$ denotes the cosmic scale factor,   conformal time $\eta$ is related to proper time $t$ by $d\eta\equiv dt/ a(t)$,  and 
\begin{equation}\label{flatspace}
    d\Sigma^2 = dr^2 + r^2 d\Omega^2
\end{equation}
represents the spatial 3-metric in comoving coordinates, with
 angular separation element $d\Omega$. Since a null path obeys
  $ d\Sigma = \pm cd\eta$,
conformal cosmological causal relationships are the same as those in flat spacetime\footnote{It is convenient to adopt units where the speed of light $c=1$, so time and length units are the same.}.

 \begin{figure*}
\begin{centering} \includegraphics[width=0.6\textwidth]{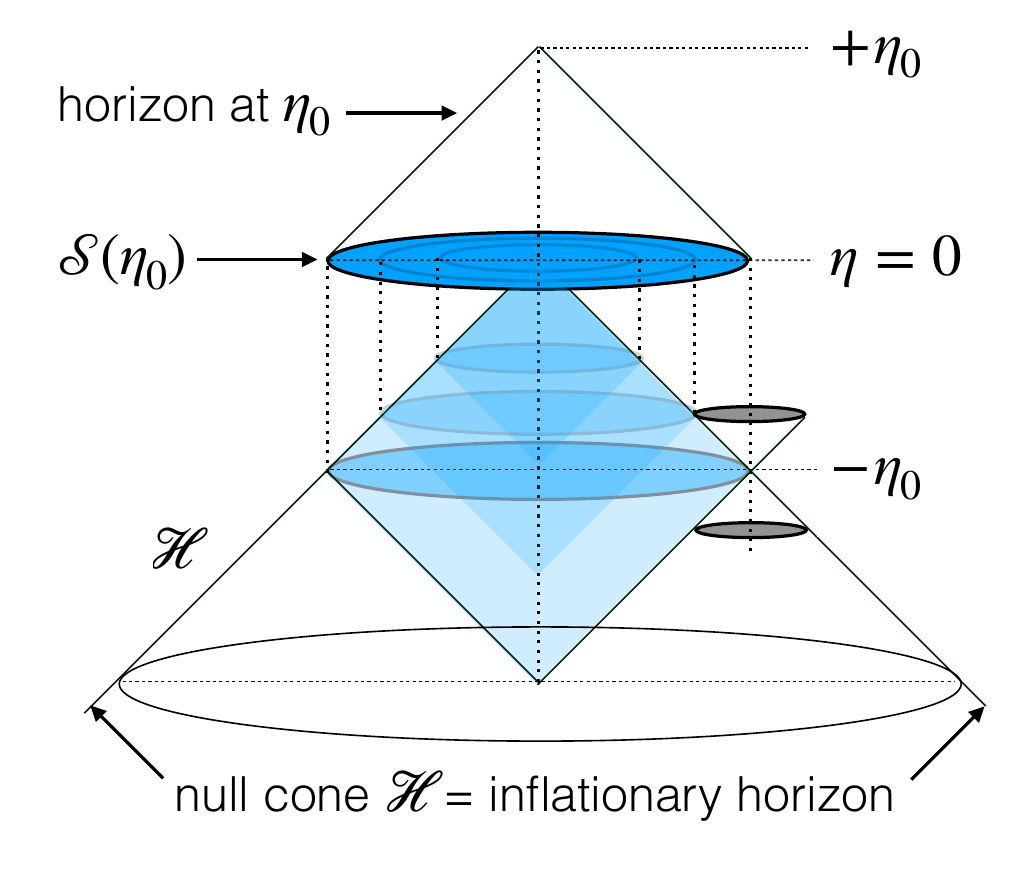}
\par\end{centering}
\protect\caption{Causal  diamonds of coherent quantum fluctuations.  Vertical axis is conformal time $\eta$, and horizontal planes represent constant-time slices of comoving space  with one spatial dimension suppressed.  Conformal time $\eta=0$ denotes the end of inflation.
Light cones show  boundaries of causal relationships with a world line. Incoming information to a world line prior to $\eta=0$ is bounded by the inflationary horizon ${\cal H}$.  Comoving spheres ${\cal S}(\eta)$ around a world line pass through its horizon  at  $-\eta$.  An observer   at a  time $\eta_0$ after inflation views the CMB last scattering surface near ${\cal S}(\eta_0)$.  Our hypothesis is that all of its  correlated perturbations are generated within a finite causal diamond that starts  at  $-2\eta_0$. Correlations in smaller volumes are generated in  coherent subsystems, the smaller nested causal diamonds as shown. Horizon-bounded causal diamonds during inflation  are all about the same physical size, which leads to a nearly scale-invariant spectrum of perturbations. Disconnected causal diamonds are separate quantum systems and form independent, uncorrelated perturbations.
 }
\label{diamonds}
\end{figure*}

Figure (\ref{diamonds}) shows   spacetime boundaries of causal entanglement.
The inflationary horizon ${\cal H}$ around every world line is the incoming spherical null surface that arrives at the end of inflation, $\eta=0$.
   A comoving spherical surface ${\cal S}(\eta)$ that passes out of contact at time $-\eta$ represents the causal boundary of correlations with its center up to that time, and comes back into view from that location after inflation at time $+\eta$. Any surface ${\cal S}(\eta)$ is the boundary of a causal diamond that starts at $-2\eta$, so a causally coherent process can in principle account for the most conspicuous correlation of cosmic structure:  nearly-perfect large-scale uniformity and isotropy, as assumed in Eq. (\ref{FLRW}). 

 
 \begin{figure*}
\begin{centering} \includegraphics[width=0.9\textwidth]{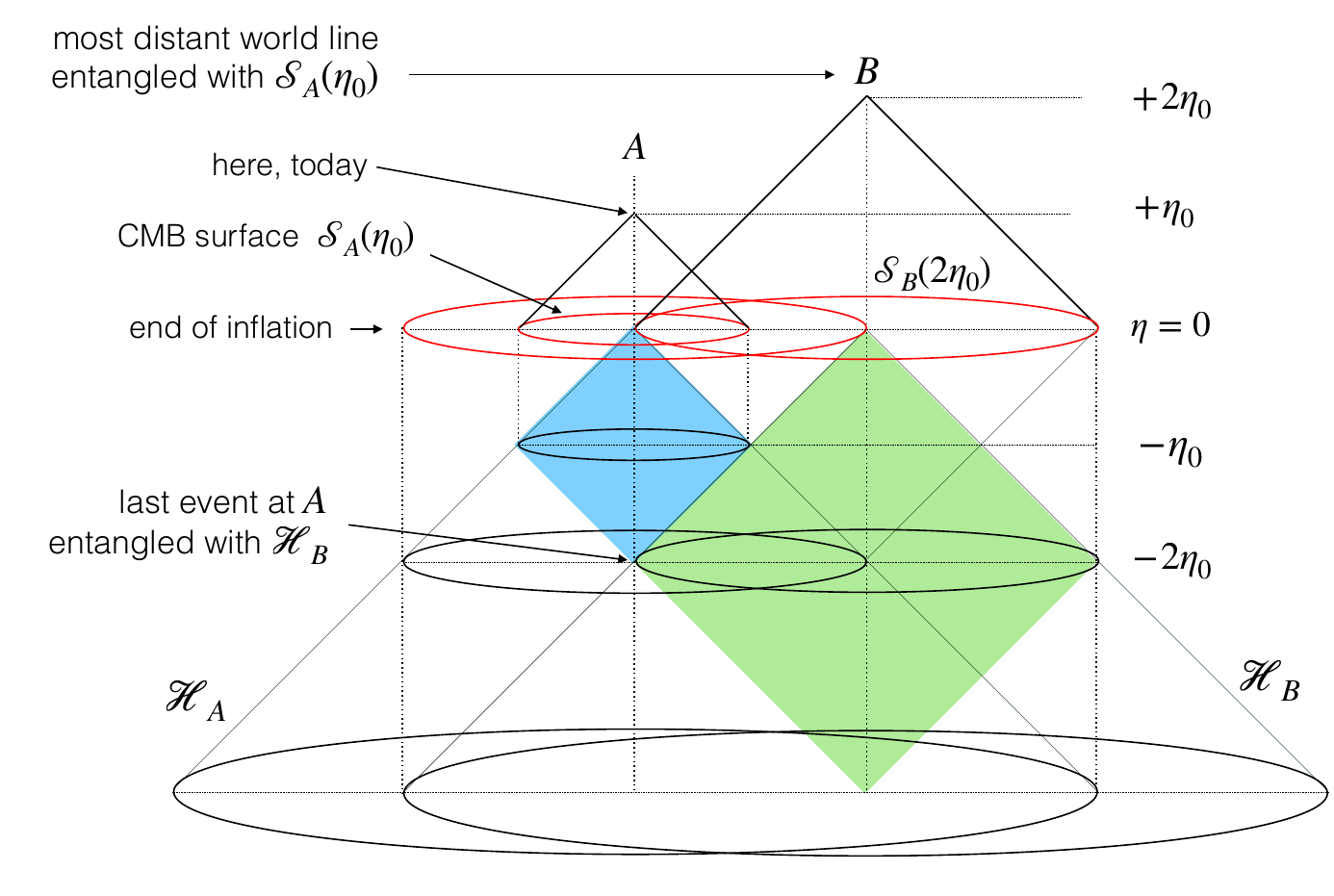}
\par\end{centering}
\protect\caption{Boundaries of causal entanglement for  two world lines and their horizons. An observer on world line  $A$  at a  time $\eta_0$ after inflation views the CMB surface on ${\cal S}_A(\eta_0)$, so its causal correlations lie within the blue shaded causal diamond that starts on $A$ at  $-2\eta_0$. The most distant world line entangled with this causal diamond is $B$, at  comoving separation $2\eta_0$.
Perturbations  within   ${\cal S}_A(\eta_0)$  are  uncorrelated with the $B$ direction outside the surface ${\cal S}_B(2\eta_0)$ of the  horizon ${\cal H}_B $. }
\label{causal}
\end{figure*}

Now consider the origin of  perturbations that depart from uniformity. These are bounded by two-way causal relationships of fluctuations around intervals on different world lines.   Consider causal diamonds around two world lines $A$ and $B$  separated by conformal time $2\eta_0$,  as shown in Figure (\ref{causal}).
Their fluctuations disentangle at time $-\eta_0$, when $A$ passes out of the horizon ${\cal H}_B$. After this,  subsequent events on $A$ initiate new  causal diamonds,  whose separate quantum fluctuations create  perturbations uncorrelated with $B$. 

\begin{figure*}[t]
\begin{centering}
\includegraphics[width=.6\linewidth]{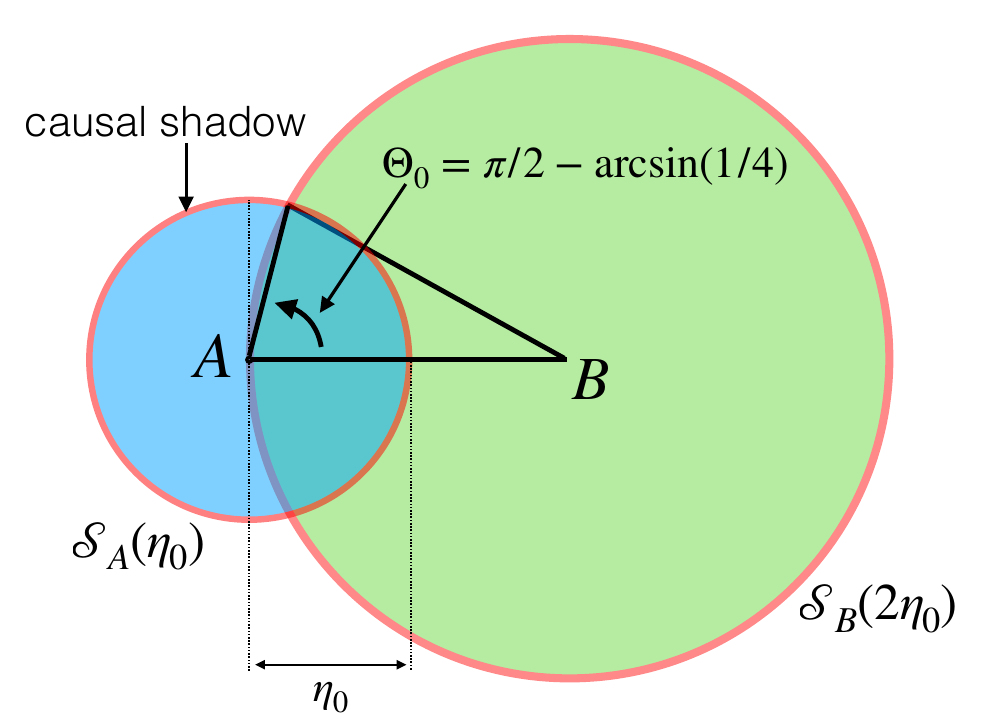}
\par\end{centering}
\protect\caption{Axial slice of spherical comoving causal diamond surfaces ${\cal S}$ at the end of inflation, showing the polar angle of the circular horizon boundary intersection in Figure (\ref{causal}). At the moment of disentanglement as shown by causal diamonds in Figure (\ref{causal}),  perturbations on  surfaces within ${\cal S}_A(\eta_0)$ have no correlation with information on horizons centered beyond ${\cal S}_B(2\eta_0)$. Closer horizons intersect ${\cal S}_A(\eta_0)$ at smaller axial angles, so no entanglement with  other world lines in the $B$ direction occurs  on ${\cal S}_A(\eta_0)$ at angular separation greater than $\Theta_0$.
\label{angle} }
\end{figure*}

Suppose we  live on $A$ at time $\eta_0$, and view the CMB on a spherical surface ${\cal S}_A(\eta_0)$.   As shown in Figure (\ref{angle}), we view the boundary of ${\cal S}_{B}(2\eta_{0})$ as a circle with a polar angular radius
\begin{equation}\label{shadowsize}
\Theta_0= \pi/2- \arcsin(1/4)\simeq 75.52^\circ.
\end{equation}
  Horizons closer than $B$ intersect ${\cal S}_A(\eta_0)$ at smaller polar angles, and more distant ones are disentangled with ${\cal S}_A(\eta_0)$, 
so $\Theta_0$ represents a maximum angular separation  for any inflationary causal entanglement in ${\cal S}_A(\eta_0)$  with  the $B$ direction. 
Since large-angle CMB correlations  are almost entirely determined by perturbations  of gravitational potential, this limit of geometrical entanglement provides a straightforward  causal explanation for    anomalously small CMB correlation at large angles: {\it primordial entanglement  of causally-coherent fluctuations does not reach large angular separations}.

\begin{figure*}[t]
\begin{centering}
\includegraphics[width=\linewidth]{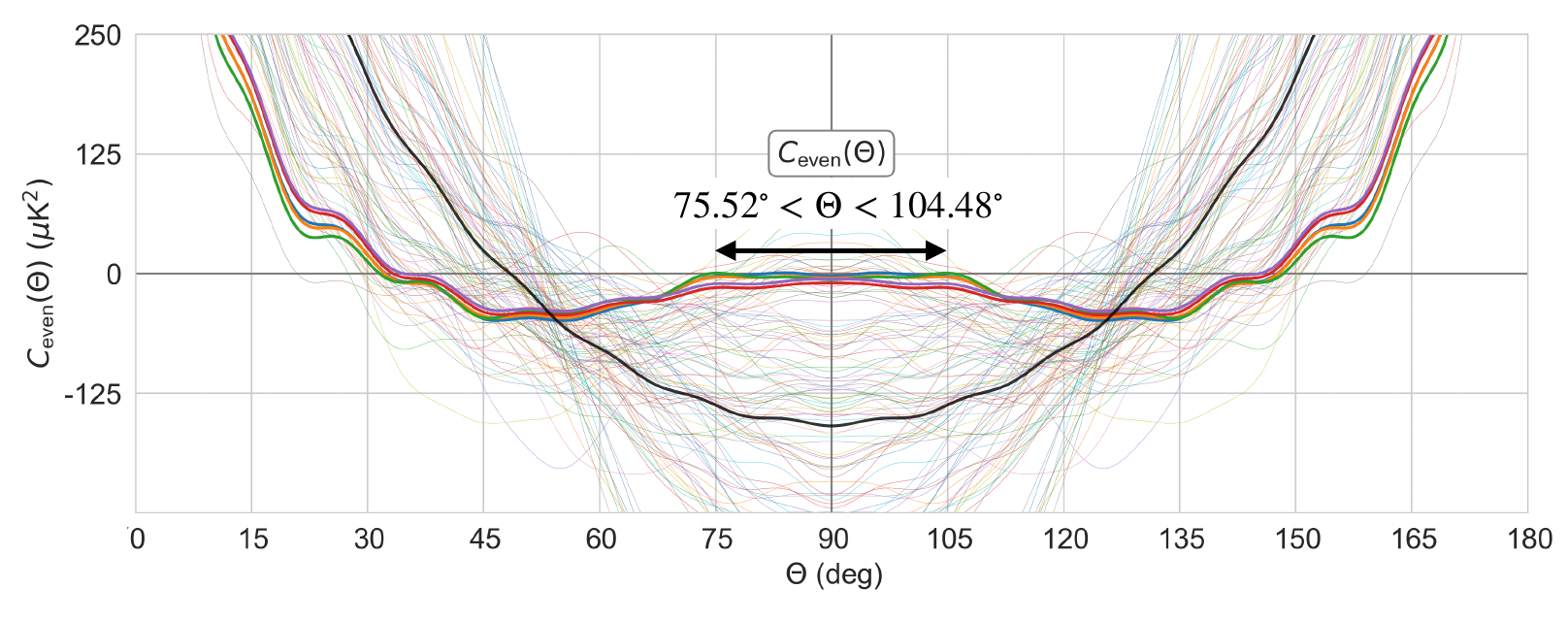}
\par\end{centering}
\protect\caption{  Even-parity correlation functions of sky maps and standard-model realizations\cite{hogan2023causal}. Bold colors show  Galaxy-subtracted all-sky maps from {\sl Planck} and {\sl WMAP}. Solid black shows the expectation of the standard model. Fine lines show standard, randomly generated sky realizations in that model.  A causal shadow  (Eq. \ref{evenshadow}) produces  even-parity correlations that  vanish over a symmetric band   $75.52^\circ<\Theta<104.48^\circ$, shown by arrows.  This symmetry is strikingly approximated by the maps, especially those from {\sl Planck}; for these, the residual correlation is  three to four orders of magnitude smaller than   most standard realizations.}
\label{evenspider}
\end{figure*}

For a precise measurement of this effect, it is useful to separate the angular correlation function into odd and even parity components \cite{hogan2023causal}.
If  correlations vanish at angular separation greater than $\Theta_0$, there is an exact symmetry or ``causal shadow'' of 
 even parity correlation   in a symmetric band  around $\Theta=90^\circ$:  
\begin{equation}\label{evenshadow}
C_{\rm even}(75.52^\circ\lesssim \Theta\lesssim 104.48^\circ)=0.
\end{equation}
The actual even-parity correlations in the CMB are shown in Figure (\ref{evenspider}).  For comparison, 100  skies are shown generated by the standard inflationary model, which is not constrained by causal coherence. 
The contrast is hard to ignore: even by eye, it is obvious that the  measured correlations lie much closer to the predicted zero (Eq. \ref{evenshadow}) than they do to typical standard realizations.  In fact, residual correlations in all the {\sl Planck} maps are three to four orders of magnitude smaller in magnitude than  typical standard realizations.  The apparent contrast is confirmed by a rank comparison: in a million trials,  the fraction of realizations  that varies as little from  zero as the  {\sl Planck} data is less than $10^{-4.3}$ to $10^{-2.8}$, depending on the map\cite{hogan2023causal}.

It is remarkable that  an exact causal symmetry,  derived in a few pages  of  geometrical reasoning from  a standard classical cosmological  metric and basic principles of causal entanglement, matches the real sky so well.
The zero correlation measured in the data--- not only at a single point, but over an extended range of angles between  precise  geometrical  boundaries calculated with no model parameters--- is hard to dismiss as a statistical anomaly. It could   be a  symmetry of a quantum process we have yet to understand.

Since   Eq. (\ref{evenshadow}) depends only on   basic principles of causal quantum relationships, why does it differ so much from the standard inflationary scenario?  We started with a hypothesis that is standard in quantum physics but not in quantum cosmology, namely, that quantum states (in this case, fluctuations of geometry in the inflationary vacuum) are coherent in causal diamonds. The quantum  field theory  used for inflation instead  assumes  coherence in comoving plane wave modes of definite direction and infinite spatial extent.  Their coherence is not bound by  information boundaries of spherical horizons, and their relict correlations extend to spacelike infinity. 
Since random realizations of  independent wave modes always lead to cosmic variance, an exact null symmetry of  correlation appears as a miraculous  conspiracy of mode phases.

Thus, causal 
 quantum coherence in causal diamonds represents a radical departure from the standard  model of gravitational quantum fluctuations.
That might not be such a bad thing:  a causally coherent theory might avoid long-wavelength theoretical difficulties noted previously\cite{CohenKaplanNelson1999,HollandsWald2004,Stamp_2015}, and perhaps even fix the famously wrong  estimated value for the cosmological constant\cite{Mackewicz2024}. 
Successful predictions of  standard cosmology are little changed, since they depend mostly on the 2-point correlation function averaged over all directions in a large 3D volume, which in turn depends mainly on the inflationary expansion rate  $H(\eta)$ and not on details of geometrical coherence or angular correlation. A detailed model might reveal small changes from standard parameter estimates, and perhaps exotic higher order signatures, such as parity violation in 3D large scale structure.

The observed angular symmetry in  primordial gravitational structure  could signify a new  geometrical symmetry in large scale quantum states of gravity, on  scales comparable to horizons, that is not included in current  theory.
A hopeful prospect is  that measured symmetries of the CMB on large scales, and perhaps other measurements of large scale cosmic structure,   could guide development of a deeper theory that  accounts for how locality and macroscopic  causal structure emerge from a quantum system.
\vfil\eject




\bibliography{grf2024.bib}

\end{document}